# WHAT TYPES OF COVID-19 CONSPIRACIES ARE POPULATED BY TWITTER BOTS?

*EMILIO FERRARA, UNIVERSITY OF SOUTHERN CALIFORNIA*
*EMAIL: emiliofe@usc.edu*

## ABSTRACT

With people moving out of physical public spaces due to containment measures to tackle the novel coronavirus (COVID-19) pandemic, online platforms become even more prominent tools to understand social discussion. Studying social media can be informative to assess how we are collectively coping with this unprecedented global crisis. However, social media platforms are also populated by bots, automated accounts that can amplify certain topics of discussion at the expense of others. In this paper, we study 43.3M English tweets about COVID-19 and provide early evidence of the use of bots to promote political conspiracies in the United States, in stark contrast with humans who focus on public health concerns.

## INTRODUCTION

At the time of this writing (mid-April 2020) the novel coronavirus (COVID-19) pandemic outbreak has already put tremendous strain on many countries' citizens, resources and economies around the world. Social distancing measures, travel bans, self-quarantines, and business closures are changing the very fabric of societies worldwide. With people forced out of the safety and comfort of their life routines, social media take centerstage, more than ever, as a mirror to global social discussions. Therefore, it is of paramount importance to determine whether online chatter reflects genuine people's conversations or otherwise may be distorted by the activity of automated accounts, often referred to as bots (a.k.a., social bots, sybil accounts, etc.). The presence of bots has been documented in the context of online political discussion [1]–[3], public health [4]–[8], civil unrest [9], stock market manipulation [10], the spread of false news [11]–[13], alongside with other tools such as troll accounts [5], [6], [14]–[17].

In this paper, we chart the landscape of Twitter chatter within the context of COVID-19 related conversation, in particular to characterize the presence and activity of bots. We leverage a large Twitter dataset that our group has been continuously collecting since January 21, 2020, when the first COVID-19 case was announced in the United States [18]. We use combinations of machine learning and manual validation to identify bots, and then use computational tools and statistical analysis to describe their behavior, in contrast with human activity, and their focal topics of discussion.

RESEARCH QUESTIONS & CONTRIBUTIONS OF THIS WORK

We hereby posit two research questions, formalized in the following, and by leveraging a large COVID-19 Twitter dataset (cf. *Methods & Data*) we provide empirical evidence as well as theoretical grounding to answer these questions:

- **Research Question 1** (RQ1): *Is there any evidence of the presence of automated accounts (bots) in the online discussion surrounding COVID-19 on Twitter? If so, what is their prevalence and volume of activity compared to that of human accounts? Do the bots exhibit any behavioral characteristic that are their prerogative, which in turn significantly differ from the behavior of human users?*
- **Research Question 2** (RQ2): *Prior work demonstrated how bots are used to push ideologies and political narratives on social media; Do we observe any pattern of preferential behavior where the bots seem to focus on fueling specific topics of discussion concerned with politics or ideology?*

In this work we provide the following contributions:

- First, we combine machine learning and human validation to identify accounts that show signatures of automation (bots) and provide a statistical characterization of their behavior, contrasted with human users, specifically in the English-speaking Twitter-sphere.
- Then, we leverage content and time-series analysis techniques to illustrate the topical focus of bots and human users, highlighting that bots appear to be used to promote conspiracy theories in the United States, in stark contrast with human users who focus on public health and welfare.

BACKGROUND & RELATED LITERATURE

There has already been a wealth of studies that looked into social media dynamics in the context of COVID-19. As of the time of this writing (mid-April 2020), the vast majority of these studies are pre-print papers that provide a timely, yet partial, characterization of online discussion and issues revolving around COVID-19 [19]–[28].

Various studies presented the concept of *social media infodemic,* i.e., the spread of questionable content and sources of information regarding the COVID-19 pandemic, as postulated by [21]. This research illustrates the problem of containing the spread of unverified information about COVID-19, showing that questionable and reliable information spreads according to similar diffusion patterns. Along this line, [27] suggests that low-quality information anticipates epidemic diffusion in various countries, with the peril of exposing those countries' population to irrational social behaviors and public health risks. Both studies account for large-scale data collection from online platforms like Facebook and Twitter but do not emphasize the importance of information manipulation on such social media. The work by Singh and colleagues also looked at the spatio-temporal dynamics of misinformation spread on Twitter, drawing a picture with similar implications as the two studies above [26].

More research is needed, as the information landscape evolves, and more scientific insights are unveiled on the clinical and medical implication of this disease, to understand what qualifies for rumors, misinformation, or disinformation campaigns. For example, information about the possible effectiveness of some treatments could be considered as rumors at a given point in time, in absence of definitive scientific consensus; yet, as more clinical evidence emerges, these may become false claims, hence classified as misinformation: one such example is Hydroxychloroquine, a known anti-malaria drug whose effectiveness in treating SARS-CoV-2 (the coronavirus causing the COVID-19 disease) remains debated at this point in time, and whose potentially lethal side-effects limit large-scale testing.

The work by Pennycook and colleagues epitomizes the seriousness of this problem by showing, with a social experiment including 1,600 participants, that subjects tend to share misinformation and false claims about COVID-19 predominantly because they are unable to determine whether the content is scientifically sound and accurate or not [22].

Other studies investigated collective attention and engagement dynamics concerning COVID-19 on Twitter. For example, [20] analyzed 1,000 unigrams (1-grams) posted on Twitter in 24 languages during early 2020, and compared them with the year prior. The authors emphasize how the global shift in attention to the COVID-19 pandemic is concentrated around January 2020, after the first wave of infections in China started to phase off, and peaked again in early March, when the United States and several other western countries started to get more heavily hit by the pandemic. Their work suggests that social media mirror offline attention dynamics, and hints at the potential implications that diminished collective attention can have on the perception of gravity of this pandemic (or lack thereof).

Various studies investigated emotional and sentiment dynamics on social media conversation pertaining COVID-19 [23]–[25]. For example, [25] annotated a corpus (N=2,500 tweets and N=2,500 longer texts) producing a ground truth dataset for emotional responses to COVID-19 content. The analysis of this UK-centric corpus suggests that issues pertaining family safety and economic stability are more systematically associated with emotional responses in longer texts, whereas tweets more commonly exhibit positive calls for solidarity.

Contrary to that, recent work based on large-scale multiplatform data collections encompassing Twitter and 4chan illustrates the endemic prevalence of hate speech, especially sinophobia, in both platforms [24]. Furthermore, the cross-platform diffusion of racial slur and targeted attacks shows how fringe platforms like 4chan are incubators of new hate narratives aimed, in the case of COVID-19, against Asian people; in mainstream social media platforms like Twitter, however, the focus is on putting blame on China for the alleged responsibility in originating the virus and inability to contain it.

On a different note, [23] looked at social media sentiment as expression of potential mental health issues associated with social isolation and other side-effects of containment measures enacted to limit the spread of COVID-19 in China. By means of online surveys to complement observational data collected from populare Chinese social media platforms, the authors suggest that social media exposure to outbreak-related content was correlated with increased odds of reporting issues associated with mental health, including depression and anxiety, across different demographics in their population.

## BACKGROUND ON COVID-19

The first cases of a novel coronavirus disease (officially named COVID-19 by the World Health Organization on February 11, 2020) were reported in Wuhan, China in late December 2019; the first fatalities were reported in early 2020. The first case in the United States was announced on January 21, 2020 [18]: our Twitter data collection aligns with that date [19].

The fast-rising infection rates and death toll led the Chinese government to quarantine the city of Wuhan on January 23, 2020.[1] During this period, other countries began reporting their first confirmed cases of the disease, and on January 30, 2020 the World Health Organization (WHO) announced a Public Health Emergency of International Concern. With virtually every country on Earth reporting cases of the disease, and infections rapidly escalating in some regions of the world, including the U.S., Europe and the middle East, WHO has subsequently upgraded COVID-19 to a pandemic.[2] On March 13, 2020 the United States government announced the state of national emergency. Our data collection's end aligns with that date. As of the time of this writing (mid-April 2020), COVID-19 has been reported in every country worldwide, leaving governments all over the globe scrambling for ways to contain the disease and lessen its adverse consequences to their people's health and economy. Infections exceed two million. Fatalities are well over a hundred thousand. There is still no scientific consensus on the effectiveness of any particular treatment; vaccines are not expected to be available to large swaths of the population for at least a year. COVID-19 has been among the trending topics of discussion on Twitter uninterruptedly since early 2020.

## BACKGROUND ON BOTS

**What is a bot**. A bot (short for *robot*) generally refers to an entity operating in a digital space that is controlled by software rather than human. Bots have been categorized according to various taxonomies [29], [30]. In this article, we use the term *bot* as a shorthand to *social bot*, a concept that refers to a social media account controlled, predominantly or completely, by computer software (a more or less sophisticated artificial intelligence), in contrast with accounts controlled by human users [31]; this is in line with the recommendations of [29] who provided the most comprehensive survey on the typologies of bots (cf., page 9: "we suggest that automated social media accounts be called social bots").

**How to create a bot**. Early social media bots, in the late 2000s, were created to tackle simple tasks, such as automatically retweeting content posted by a set of sources or finding and posting news from the Web. Today, the capabilities of bots have significantly improved: bots rely on the fast-paced advancements of Artificial Intelligence, especially in the area of natural language generation, and use pre-trained multilingual models like OpenAI's GPT-2 [32] to generate human-like content. This framework allows the creation of bots that generate genuine-looking short texts on platforms like Twitter, making it harder to distinguish between human and automated accounts [33].

The barriers to bot creation and deployment, as well as the required resources to create large bot networks, have also significantly decreased: for example, it is now possible to rely upon bot-as-a-service (BaaS), to create and distribute large-scale bot networks using pre-existing capabilities provided by

---

[1] https://www.nytimes.com/article/coronavirus-timeline.html
[2] https://www.who.int/emergencies/diseases/novel-coronavirus-2019/events-as-they-happen

This paper has been peer reviewed and accepted for publication in First Monday companies like *ChatBots.io*, and run them in cloud infrastructures like *Amazon Web Services* or *Heroku*, to make their detection more challenging [34].

**Open source Twitter bots.** A recent survey discusses readily-available Twitter bot-making tools [35]: the authors provide an extensive overview of open-source GitHub repositories and describe how prevalent different automation capabilities, such as tweeting or resharing, are across these tools.

According to [35], whose survey focused exclusively on repositories for Twitter bots developed in Python, there are hundreds of such open-source tools readily available for deployment. The authors studied 60 such bot-making tools and enumerated the most common capabilities. Typical automated features of such bots include:

(1) searching users, trends, and keywords;
(2) following users, trends, and keywords;
(3) liking content, based on users, trends, and keywords;
(4) tweeting and mentioning users and keywords, based on AI-generated content, fixed-templated content, or cloned-content from other users;
(5) retweeting users and trending content, and mass tweeting;
(6) talking to (replying) other users, based on AI-generated content, templated content, or cloned-content from other users; finally,
(7) pausing activity to mimic human circadian cycles and bursty behaviors, as well as to satisfy API constraints, and to avoid suspension.

According to [35], these features can enable bots to carry out various forms of abuse including: denigrate and smear, deceive and make false allegations, spread misinformation and spam, and finally clone users and mimic human interests. We refer the interested readers to the excellent survey by [35] for further details.

**Bot behavior research**. Numerous studies have been devoted to characterize the behavior of bots [36]–[38] and the ethical issues pertaining their use [39]. E.g.: According to [30], [40], [41], the behavior of political bots can be categorized with respect to their goals:

- Manufacturing consensus, to enhance the perception of popularity or influence of an entity (political actor, party, organization, etc.);
- Bolstering opinions and voices, to amplify the platform and audience that an entity will receive;
- Cementing polarization, by increasing the inflammatory or divisive nature of an issue or agenda;
- Increasing chaos and confusion, by posting inaccurate information, disinformation, and rumors;
- For algorithmic manipulation, to trick recommendation and ranking systems used by social media platforms, and give higher visibility to certain actors, viewpoints, or campaigns.

According to work by [5], [42], similar conclusions can be drawn for bots active in public health discussions, with the intent to spread claims contrary to scientific evidence.

**How to detect bots**. Researchers in cyber-security have highlighted first some potential challenges associated with the detection of bots [43]–[47]. Historically, however, bot detection techniques have been pioneered by groups at Indiana University, University of Southern California, and University of Maryland, in the context of a program sponsored by DARPA (the U.S. *Defense Advanced Research Projects Agency*) aimed at detecting bots used for anti-science misinformation [4].

More recently, large bot networks (botnets) have been discovered on Twitter by various research groups [43], [44], [46], [48].

The literature on bot detection has become very extensive [31], [36], [49]–[51].

In [31], we proposed a simple taxonomy to divide bot detection approaches into three classes: (1) systems based on social network information; (2) systems based on crowd-sourcing and the leveraging of human intelligence; (3) machine learning methods based on the identification of highly-predictive features that discriminate between bots and humans. Other recent surveys propose complementary or alternative taxonomies that are worth considering as well [36], [50], [51]

Some openly accessible tools exist to detect bots on platforms like Twitter:

(1) *Botometer*[3], also used here, is a bot-detection tool developed at Indiana University [52];
(2) *BotSlayer*[4] is an application to detect and track potential manipulation of information on Twitter;
(3) the *Bot Repository*[5] is a centralized database to share annotated datasets of Twitter bots.

Finally, various models have been proposed to detect bots using sophisticated machine learning techniques, such as:

(1) Deep learning [53],
(2) Anomaly detection [54]–[57],
(3) Time series analysis [38], [58], [59].

Due to the continuously evolving nature of bots, and the challenges that that poses for detection, in this article we will focus on studying the top and bottom end of the bot score distribution, rather than carrying out a binary classification of accounts into bots and humans.

This avoids the conundrum of dealing with borderline cases for which detection can be inaccurate, and conversely to focus on accounts that exhibit clear human or bot traits. Furthermore, the results will be manually validated for accuracy.

---

[3] Botometer: https://botometer.iuni.iu.edu/
[4] BotSlayer: https://osome.iuni.iu.edu/tools/botslayer/
[5] Bot Repository: https://botometer.iuni.iu.edu/bot-repository/

# RQ1: CHARACTERIZING BOT AND HUMAN BEHAVIOR

In this section, we address **RQ1** with a statistical characterization of bot behavior in our data.

## BOT SCORE RANK DISTRIBUTION

Bot detection is hard. Bot-making tools continuously evolve, and the capabilities of bots improve while available bot detection techniques catch up. For this reason, it becomes increasingly harder to classify users "in the wild" preserving high degrees of accuracy across the whole spectrum of human-to-bot likeness.

For such a reason, in this study we focus on the top and bottom end of the bot score rank distribution, and isolate accounts in the top decile (i.e., top $10^{th}$ percentile of the bot score distribution) and flag them as *high bot score accounts*; conversely, we isolate users in the bottom decile (i.e., bottom $10^{th}$ percentile of the distribution), and refer to them as *low bot score accounts*. We will only draw distinctions at the aggregate level between these two groups, without making any further inference, either binary or probabilistic classification, of the nature of any given account.

Similar approaches have been developed in the domain of information warfare detection: one such example, is the CUT (Commenting User Typology) Framework, which was proposed to classify comment abuse, where users are divided into quadrants of behaviors alongside various dimensions of interest [60].

The idea that there exists a continuum of behavior along the dimensions of social media engagement is also in line with two other theories, namely the notions of "super-participation" [61] and that of "dynamical classes of behavior" [62]. In our case, we adopt the same idea to segmenting the dimension of human-to-bot likeness and dividing the continuous spectrum of the bot score dimension into percentiles, then focus on the upper and lower deciles of the distribution, which yields two distinct groups obtained with respect to the behavior of interest.

In **Table 3**, we show the percentile rank distribution of bot scores and average values of a subset of selected user activity features, namely (i) the total number of tweets posted by each users, (ii) the proportion of COVID-19 related tweets observed in our data, (iii) and the account age, measured as the number of days elapsed between the creation of the account and their first COVID-19 tweet in our data. The distribution portrays such aggregate statistics every fifth percentile of the bot scores.

Some striking patterns emerge. As the bot scores increases, the number of total tweets posted by users on average decreases. For example, users in the bottom $5^{th}$ percentile (0.05) have posted on average over 15 thousand tweets. Conversely, accounts in the top $5^{th}$ percentile, have posted on average only about 1,600 total tweets. A similar pattern emerges with account age. Accounts in the bottom end of the bot score distribution have been active on average for almost three thousand days (or 8 years!) as opposed to accounts with the higher bot scores, whose average age is less than three years.

Table 1: Rank distribution of bot scores and account activity average metrics, along with suspended and verified statistics.

| Percentile | Bot Score | Total Tweets | COVID-19 Ratio % | Account Age | Suspended % | Verified % |
|---|---|---|---|---|---|---|
| 0.05 | 0.03 | 15312 | 0.03 | 2909 | 2.34 | 1.9 |
| 0.1 | 0.04 | 13287 | 0.04 | 2919 | 2.12 | 1.97 |
| 0.15 | 0.04 | 10986.5 | 0.06 | 2864 | 2.11 | 1.73 |
| 0.2 | 0.05 | 9223 | 0.07 | 2742 | 2.05 | 1.48 |
| 0.25 | 0.05 | 8009 | 0.09 | 2585 | 2.11 | 1.34 |
| 0.3 | 0.06 | 7151 | 0.1 | 2418 | 2.29 | 1.11 |
| 0.35 | 0.07 | 6592.5 | 0.12 | 2232 | 2.42 | 0.94 |
| 0.4 | 0.08 | 6036.5 | 0.13 | 2054 | 2.56 | 0.8 |
| 0.45 | 0.09 | 5566 | 0.15 | 1827 | 2.61 | 0.74 |
| 0.5 | 0.1 | 5122 | 0.17 | 1551 | 2.9 | 0.59 |
| 0.55 | 0.12 | 4621 | 0.19 | 1399 | 3.09 | 0.54 |
| 0.6 | 0.14 | 4375 | 0.21 | 1237 | 3.3 | 0.44 |
| 0.65 | 0.16 | 4016 | 0.24 | 1176 | 3.42 | 0.37 |
| 0.7 | 0.19 | 3752 | 0.27 | 1134 | 3.42 | 0.34 |
| 0.75 | 0.22 | 3601 | 0.3 | 1136 | 3.47 | 0.41 |
| 0.8 | 0.27 | 3416 | 0.33 | 1187 | 3.21 | 0.29 |
| 0.85 | 0.34 | 3510 | 0.35 | 1314 | 2.74 | 0.24 |
| 0.9 | 0.44 | 4178.5 | 0.31 | 1503 | 2.48 | 0.17 |
| 0.95 | 0.59 | 5112 | 0.27 | 1609 | 2.07 | 0.15 |
| 1 | 0.79 | 1626.5 | 0.81 | 1184.5 | 2.24 | 0.06 |

However, the trend is reversed when looking at the fraction of COVID-19 related tweets: accounts with higher bot scores post significantly more COVID-19 tweets than those in the lower end of the distribution. In fact, for accounts in the top 5th percentile of the bot score distribution, the ratio of COVID-19 tweets to their total is 0.81%, whereas for the bottom 5th percentile this ratio is 0.03%. In other words, accounts with the highest bot scores post about 27 times more about COVID-19 than those with the lowest bot scores.

Suspensions across the bot score spectrum vary from about 2% to approximately 3.5%, with accounts having bot scores in the 60-80 percentile being more likely to get suspended. Concluding, only 81 accounts (0.1%) with bot scores in the top decile are verified. Accounts in the bottom decile are on average twenty times more likely to be verified (avg. ~2%) than the top decile (avg. 0.1%).

Overall, the insights drawn from the bot score rank distribution analysis suggest that an investigation to characterize the behavior of suspicious accounts with high bot scores is warranted.

BOT SCORE DISTRIBUTION VALIDATION

To provide additional insights in the bot score distribution, we leverage the annotations of verified and suspended users. In **Figure 1**, we illustrate the histogram of bot scores for verified and suspended accounts in our dataset. The two distributions are statistically very significantly different (Mann-Whitney rank test, p-value<0.001).

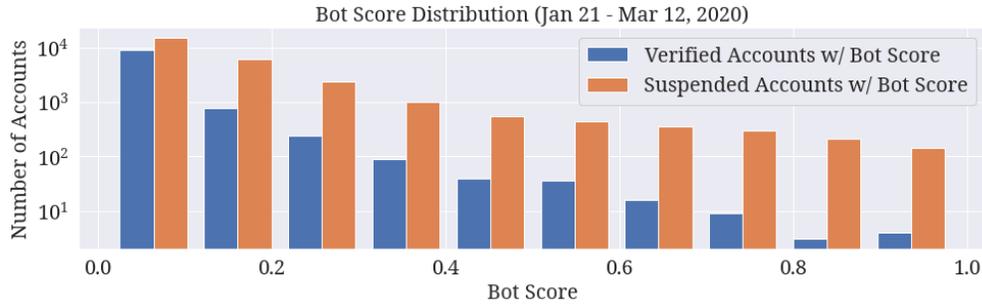

Figure 1: Distribution of bot scores for verified and suspended accounts in our dataset.

Whereas approximately 90% of the verified accounts have bot scores lesser than 0.1, the bot score of suspended accounts is much more broadly distributed, with approximately half of the suspended accounts exhibiting scores higher than 0.1. Whereas in the lower end of the distribution there are both suspended and verified accounts – which is to be expected, since accounts can be suspended for various reasons, not just for being automated – the upper end of the distribution does not contain almost any verified user, but it exhibits hundreds of suspended accounts. This suggests that there is a correlation between account suspension and increased bot likeness.

## HIGH AND LOW BOT SCORE USER ANALYSIS

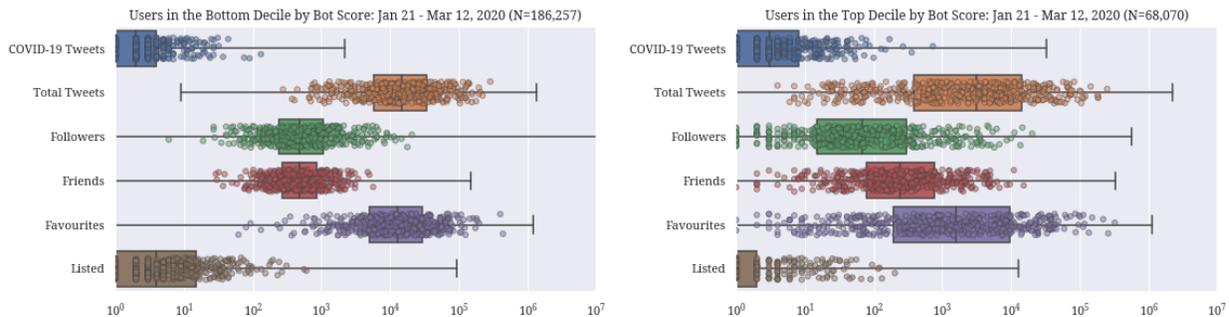

Figure 2: Distributions of user activity features for low bot score users (left) versus high bot score accounts (right). Whiskers represent the inter-quartile range of the distributions and the jitter is random samples from the underline distributions.

In line with recent analysis by [63], as well as our prior work [2], we report six basic account meta-data features that are known to carry predictive power in the differentiation between bot and human users, namely (i) topical tweets (in this case, COVID-19 related tweet count), (ii) total number of tweets, (iii) number of followers, (iv) number of friends, (v) number of favorited tweets, and (vi) number of times the account was added to a list by other users.

In **Figure 2**, we show the strip plots of the distributions for all features, for accounts in the bottom decile of the bot score distribution (left plot) as well as for accounts in the top decile (right plot). The strip plots convey all observations alongside samples of the underlying distribution data, displayed as jitter over the bot plot that is randomly sampled from the underlying distribution. Visual comparison of the two plots illustrates immediately a striking difference in the distributions: in line with **Table 3**, high bot-score accounts have significantly more COVID-19 tweets but fewer total tweets, they have significantly fewer followers (one order of magnitude difference), fewer of their tweets were favorited by other users (one order of magnitude difference) and they were added to fewer public lists. These differences are confirmed by statistical analysis: the feature-pairwise Mann-Whitney rank tests are all strongly significant, all p-values<0.001.

### *AGE AND PROVENANCE ANALYSIS*

Our final investigation in the characteristics of high and low bot score accounts centers around age and provenance of the user accounts. The analysis above suggested than on average user accounts with higher bot scores in our COVID-19 dataset also exhibit shorter account age. Account age and prevalence of activity related to COVID-19 appear to be very strongly correlated features.

To further investigate this relation, in **Figure 3** we show the distributions of account age of user accounts at the time when they joined the COVID-19 discussion. The histogram portrays two very distinct stories for accounts in the top and bottom deciles of the bot score distribution: the former appear to be joining the COVID-19 in the early days since their creation: in fact, the average amount of time elapsed between account creation and first COVID-19 post for high bot score users is less than 100 days. In other words, the vast majority of high bot score accounts have been created relatively in proximity to the emergence of the COVID-19 outbreak and jump on this discussion with high intensity shortly after their creation. For example, over 80,000 high bot score accounts have been created between 50 and 100 days prior to their first COVID-19 tweet. Conversely, it is apparent that accounts in the bottom decile of the bot score distribution have been created significantly prior to the events.

The second aspect that we evaluate is the provenance of these accounts. The attribution of account provenance is a well-known challenging task, because the Twitter API does not provide crucial forensic information that would be required for exact provenance attribution, such as the IP address of the machine or VPN server an account connects from. However, in lieu of such information, the best proxy at our disposal is the ability to reconstruct the server and data center that dispatched each tweet. There are two data centers, namely DC10 and DC11, that dispatch tweets, and 30 servers associated with these data centers. A simple language analysis of the tweets originating from the two data centers clearly suggests that DC10 is used to dispatch tweets originating in Asia, South-East Asia, Russia, and the Middle East. Conversely, DC11 dispatches tweets originating predominantly from Europe and the Americas.

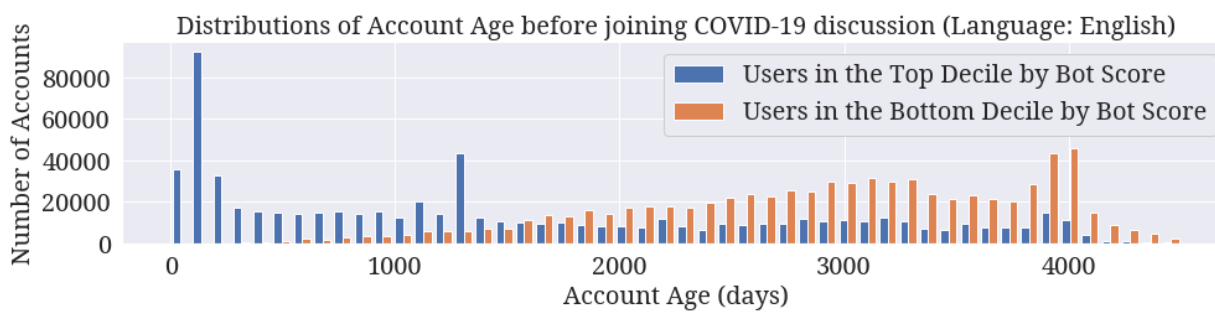

Figure 3: Age distribution of the accounts in the top and bottom decile of the bot score distribution at the time of joining the COVID-19 Twitter discussion.

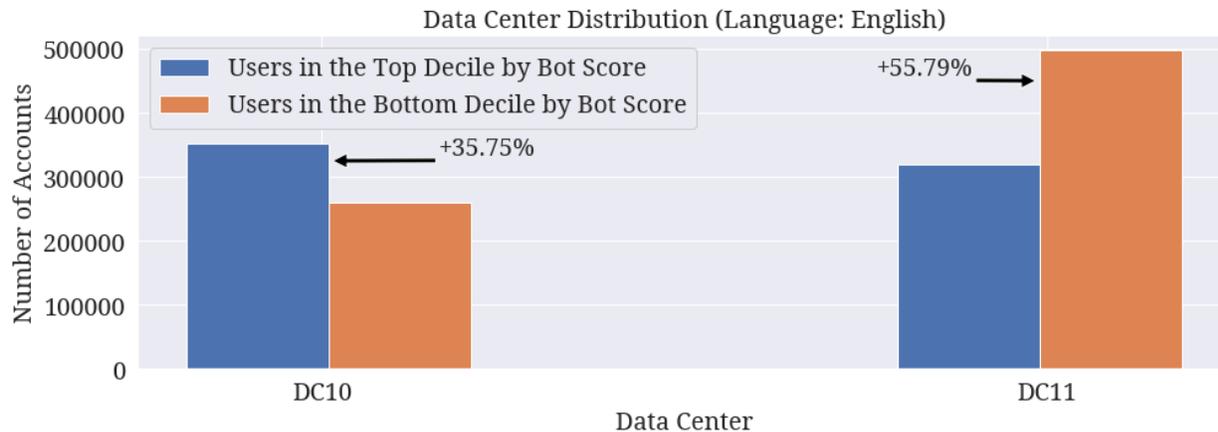

Figure 4: Distributions of the data center provenance of accounts with high and low bot scores.

**Figure 4** illustrates remarkable differences in the data center connectivity patterns between accounts in the top and bottom deciles by bot scores. In particular, it appears evident that DC10 (the data center that serves the Eastern world) dispatches over 35% more tweets originating from high bot score accounts (blue bars) than from low bot score ones (orange bars). The opposite patterns appear to be true for DC11 (the data center serving the Western world): DC11 dispatches over 55% more tweets from low bot score users than from top score ones. We can only speculate about the origin of such difference, in absence of the investigative tools necessary to get a definitive answer: prior investigations carried out by Twitter unveiled the systematic presence of information operations based in countries such as Russia, Iran, North Korea, and China [64], and this pattern seems to emerge again for COVID-19 discussion [65]. We speculate that a fraction of these may be carried out using bots, which in turn leave a digital trail associated with the data center used to dispatch tweets.

## RQ2: CHARACTERIZING BOT FUELED CAMPAIGNS

Data-driven computational sensemaking can fail when using human communication data [66]. Common reasons include population and sampling bias, platform-design bias, distortion of human and nonhuman behavior, multiple hypotheses and comparisons testing, and more, which can hamper computational models and bias results [67]. An hybrid approach based on combining computational analysis and human validation has been called for in order to make sense of massive-scale communication datasets [68], [69].

**Content sensemaking**. In this section, we address **RQ2** using content and timeseries analysis techniques, in combination with manual content analysis and validation, to characterize bot fueled campaigns. We use two distinct strategies, namely keywords and hashtag analysis. Specifically, keywords are obtained by means of n-gram analysis (an n-gram is simply a sequence of *n* words). Looking at frequently occurring keywords and hashtags represents a common approach to surface trends of interest in social media datasets [70]. Manual validation for both strategies allows us to corroborate the findings derived from content-based computational analysis and provide interpretations aimed at answering RQ2.

## SURFACING CHARACTERISTIC CONTENT PRODUCED BY BOTS AND HUMANS

The goal of the following analysis is to identify patterns of information production and consumption that are characteristic of the high (resp., low) bot score populations. To this aim, we will first isolate all tweets produced by these two groups and carry out a comparative analysis in the adoption patterns of keywords and hashtags for sensemaking purposes. Then we will surface and discuss the expected differences that likely-automated accounts exhibit with respect to likely-human users. Next, we detail the preprocessing steps taken to curate the textual content (tweets) produced and consumed by the two groups.

**Preprocessing**. The first step was to isolate tweets in English language. The Twitter API provides the estimation of the language, which we leveraged to select a subset of approximately 43.3M tweets. Out of this set, we isolated the tweets produced by the high (resp., low) bot score users (i.e., the users in the top and bottom deciles of the bot score distributions). This produced 671,774 tweets total tweets for the *high bot score accounts*, and 756,940 tweets for the *low bot score users.* Starting from these tweets, we will extract the characteristic hashtags and n-grams preferentially adopted by the two groups of accounts.

**Tweet type disaggregation**. On twitter, there are four modalities of posting: (i) original tweets; (ii) reply tweets; (iii) quoted tweets; (iv) retweets. Each of these mechanisms is used for different purposes. An original tweet is posted any time a user composes a new tweet from scratch. A reply is an answer to another tweet, typically posted by another user, albeit it is possible to reply to one's own tweets. A quote embeds another tweet and adds original text typically as a commentary; it's once again possible to quote one's own tweets, however typically a quote embeds tweets posted by other users. Finally, a retweet is a one-click operation that allows to reshare on one's timeline another tweet that will appear without any modifications or commentary (again, typically posted by another user, despite it's possible to retweet one's own tweets). In the following analysis, we will disaggregate according to these four communication mechanisms, since they have different aims, and can also be abused in specific ways. By means of this disaggregation, we obtain the following subsets of tweets:

(i)  50,483 and 83,342 original tweets posted by high and low bot score accounts, respectively;
(ii)  10,852 and 50,756 reply tweets posted by high and low bot score accounts, respectively;
(iii)  70,432 and 153,304 quote tweets posted by high and low bot score accounts, respectively;
(iv)  540,007 and 468,539 retweets posted by high and low bot score accounts, respectively.

It's worth observing how *high bot score accounts* appear to disproportionally predilect the adoption of retweets (which is a one-click, or if you wish a one-line-of-code operation) whereas *low bot score users* tend to produce significantly more original, reply, and quoted tweets. For example, *low bot score users* produce nearly five times more reply tweets, and more than twice quoted tweets, than *high score ones*.

**N-gram extraction**. We will carry out a systematic n-gram analysis to surface common sub-sentences that tend to occur frequently in the tweets. We carried out n-gram extraction for n=1, 2, and 3. For n=1 we obtain 1-grams, a.k.a. unigrams. For n=2 we obtain 2-grams, a.k.a. bigrams. Finally, for n=3 we obtain 3-grams, a.k.a. trigrams. In the following, we discuss the trigram analysis which provides the most interpretable results.

Each tweet in the corpus is processed according to the following cleaning protocol. First, the tweet text is lower-cased. Then, link (URLs) are removed, alongside user mentions (username of other Twitter accounts which are preceded by the "@" symbol), and hashtags (terms preceded by the "#" symbol) – a hashtag analysis is indeed carried out separately. Special characters are also removed, to clean the tweet text from non-linguistic symbols such as ampersands, etc. Finally, stop-words, common English-language terms that include short function words, as well as non-lexical words, are also removed using the "nltk" Python library. Finally, for each n-gram we check that at least one word is longer than 4 characters, to remove common n-grams typical of online slang, such as "lol" or "ah ah", etc. The n-grams extracted by this process will be analyzed in the next sections.

**Hashtag extraction**. The Twitter API provides the list of hashtags included in each tweet, which we leverage to extract all hashtags from all 671,774 tweets in the *high bot score accounts* group (resp., 756,940 tweets in the *low bot score users* group). For each tweet, we remove hashtags that contain the keywords that we used to seed the data collection, which are listed in **Table 1**. For example, if a tweet contains two hashtags, e.g., #coronavirus and #ncov2019, the former would be removed because it contains the keyword "corona" that we used to seed the data collection. The removal of hashtags containing seed keywords make surfacing other interesting hashtags easier. We decided to avoid any other preprocessing step, such as removing hashtags shorter than some number of characters, or hashtags that do not appear above a certain threshold, in order to avoid skewing the results in any way. This means that at times we will observe hashtags that exhibit low volume, for example hashtags that contain a misspell, e.g., "coronoavirus" can be observed.

## CONTENT ANALYSIS: CONSPIRACIES & POLITICAL PROPAGANDA

In the following, we present the core findings that our analysis highlighted. We discuss how a subset of *high bot score accounts* appear to be engaged in the spread of conspiracies and political propaganda, in stark contrast with the comparison group of *low bot score users* that is instead engaged in the discussion of public health concerns. We will also provide examples of both trends later in the *Validation* section.

The spread of conspiracies on online social media is a well-established issue. Numerous studies have been devoted to understand, for example, how unscientific claims circulate online [71]–[74], or how conspiratorial narratives are constructed online [75]–[77], especially in the context of political ideology [2], [78]. Our analysis highlights the emergence of similar issues in the context of the conversation related to COVID-19.

**Characteristic n-gram analysis**. In **Figure 5**, for illustration purposes, we show the timeseries of the top 4 characteristic trigrams produced by three populations: (A) bots fueling conspiracy theories, (B) bots posting COVID-19 news; (C) human users (i.e., low bot score accounts). An n-gram is considered characteristic in this analysis if it appears in the top 10 of a group (e.g., the *bots*) but does not appear in the top 10 of the other (e.g., the *human users*). This allows us to surface the most popular characteristic semantic trends in each group. Each timeseries in **Figure 5** shows the daily volume of tweets containing a given trigram in that population exclusively. It is important to underscore, again, that this is the prevalence of the n-grams in each group, and not in the whole Twitter population – this is in order to give a perspective on the relative volume of content produced by each group relative to each other.

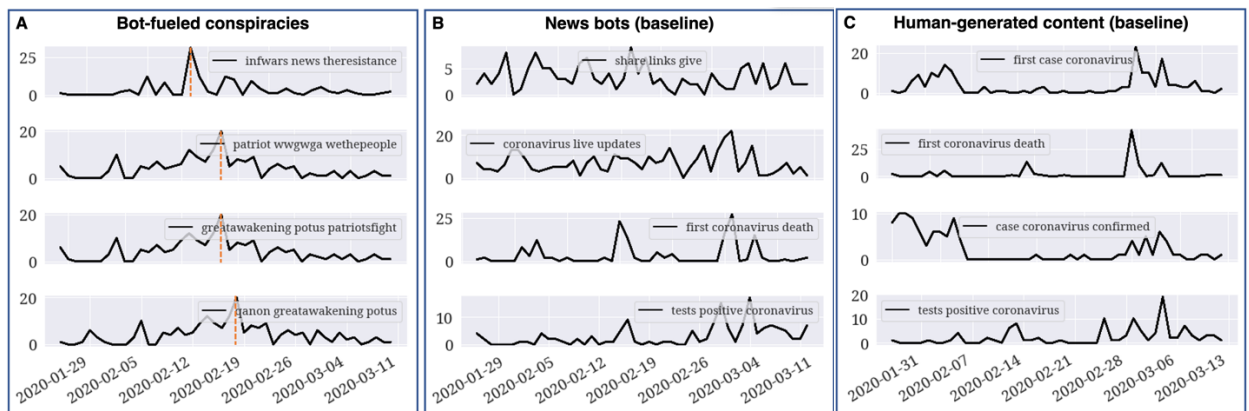

Figure 5: (A) Top 4 characteristic bot-generated conspiracy-related trigrams (first peaks are highlighted by orange lines); (B) Top 4 trigrams from news bots (baseline); (C) Top 4 human-generated trigrams.

## FINDINGS

**Figure 5A** suggests that there exists a set of *bots* that is predominantly posting conspiratorial content of political nature in the context of the COVID-19 discussion. **Figure 5B** shows the trigrams discussed by news-posting bots, another type of bot that is frequently observed in social media discussion pertaining real-world events, according to [29], [41]. Both panels A and B show content posted by *high bot score accounts*. **Figure 5C** shows trigrams posted by likely human users (*low bot score accounts*). Similar conclusions hold if one looks at a broader set of hashtags, not just the top 4 of each group.

By comparing the volume of bot-fueled conspiracies (**Figure 5A**) against the amount of discussion by human users (**Figure 5C**), we notice that the activity associated with these narratives driven by *bots* is in all comparable with that of *low bot score users* concerned with public health risks.

Various alt-right popular narratives can be isolated which are aimed at pushing divisive political ideology.

The top characteristic bot-driven trigrams in this analysis contain keywords such as *QAnon*, which according to Wikipedia,[6] is "is a far-right conspiracy theory detailing a supposed secret plot by an alleged "deep state" against U.S. President Donald Trump and his supporters". *QAnon* has been extensively adopted by alt-right activists to foster participatory advocacy on social media [79], but it has also been abused by the Russian Internet Research Agency (IRA) to push conspiratorial and divisive narratives [80], [81]. *QAnon* appears alongside with other known alt-right terms, e.g., *GreatAwakening*, *InfOwars*, WWGWGA (Where We Go One We Go All), *WeThePeople*, and *PatriotsFights* [82].

In **Figure 5A**, the first peak of each trigram timeseries is highlighted in orange. A trend emerges: these conspiratorial n-grams appear to all trend around the same time, namely during the second and third week of February 2020. This may be a clue to an orchestrated effort to push this propaganda in a coordinated manner, which is in line with recent findings on group-coordinated disinformation efforts [83], [84]. In detail, it appears that a first wave of bot-driven campaigns pushing content from hyperpartizan news site Infowars, peaked on February 12 (cf. **Figure 5A**, top timeseries), and paved the way to a second wave of bot-fueled trends including the *QAnon*, *GreatAwakening* and *WWGWGA* narratives that appeared in the following days (**Fig 5A**, other timeseries).

---

[6] https://en.wikipedia.org/wiki/QAnon

The news-bots serve as a baseline to illustrate the typical time-series patterns that are observed in non-conspiratorial bots. Bursts of activity are prerogative of coordinated operations, and differ from other news-bots' trends, such as "share link" and "coronavirus live updates" bots that post news about COVID-19 automatically and therefore do not exhibit such bursts of activity (cf. **Figure 5B**).

For what pertains humans and their activity, it is worth noting how most of the activity seems to be associated with two time periods, namely the early observation phase (the end of January and early February) and the final period (cf. **Figure 5C**). This is in line with what has been already observed by [20]: English-speaking human users seemed collectively focused on COVID-19 predominantly when it arrived in the US, and after the first fatalities started to occur in the US.

To the contrary, bot-fueled conspiracies appeared in the between these two spikes and filled the gap shifting the focus from public health to political ideology for a brief period. The peaks of these conspiracy trends (highlighted in orange in **Figure 5A**) exhibit the sequential pattern earlier-described, where bots pushing Infowars content lead the way to the subsequent trending of other alt-right keywords.

The activity of conspiracy-fueling bots appears to dissipate toward the end of our observation window (early March). This type of bot behavior has been documented already in the past, and is commonly related to as "trend hijacking" [85], [86]: bots appear to ride the wave of popularity of a given topic, in this case the trending COVID-19 discussion, to inject a deliberate message or narrative in order to amplify its visibility.

**Characteristic hashtag analysis.** In **Figure 6**, we illustrate the timeseries of top 4 characteristic hashtags characterizing (**Fig 6A**) the conspiracy-fueling bots, (**Fig 6B**) the news bots; and (**Fig 6C**) the human users, in original tweets. Both **Figures 6A** and **6B** focus on *high bot score accounts,* whereas panel C captures *low bot score users*. Similar to for the n-gram analysis, hashtags are deemed characteristic to a group if they appear in the top 10 of that group and do not appear in the top 10 of the other. We note how the top 10 of characteristic hashtags for the *human users* contains, again, predominantly hashtags that are associated with the public health aspects of the COVID-19 pandemic (cf. **Figure 6C**). Further content analysis shows that the most common characteristic hashtags include news-related terms like #breaking, mentions of influential actors (e.g., #trump), organizations (e.g., #who), and countries (e.g., #iran), alongside with the COVID-19 hashtags used to characterize the disease-related topic, including #2019ncov, #ncov, and #ncov2019.

On the other hand, for *high bot score accounts*, we observe once again a picture compatible with the n-gram analysis discussed above. Alongside with news-related hashtags such as #news and #smartnews (cf. **Figure 6B**), we observe some alt-right hashtags such as #qanon and #greatawakening (cf. **Figure 6A**). The peaks of conspiracy-fueling bots (highlighted in orange in **Figure 6A**) once again occur in the middle of our observation period, whereas news-posting bots appear active throughout the whole period, and human users exhibit activity toward the early and late observation period, namely late January and early March.

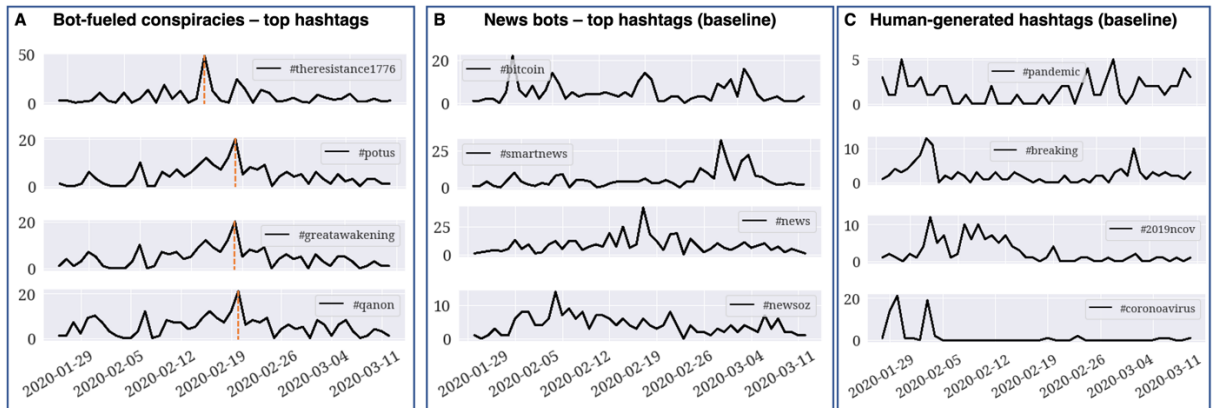

Figure 6: (A) Top 4 characteristic bot-generated conspiracy-related hashtags (first peaks are highlighted by orange lines); (B) Top 4 hashtags from news bots (baseline); (C) Top 4 human-generated hashtags.

## VALIDATION

Taken together, the n-gram and hashtag analyses paint a picture suggestive of the fact that *high bot score accounts* are injecting content with conspiratorial narratives charged with alt-right ideology. These hypotheses are further validated with a process of manual verification and coding.

The first step in our validation was to determine how many verified accounts in the *high bot score* population posted about conspiracies and political propaganda. For sake of illustration, we discuss the results for the trigram analysis discussed above. This validation step determined that only 9 verified accounts posted content (out of 1803 total, or 0.05%).

Conversely, for *low bot score users*, the number of verified accounts was 215 (out of 1037 total, or 20.7%). In other words, in the trigram analysis above, *low bot score users* were 414 times more likely to be verified than *high bot score accounts.* The former population consisted predominantly of established social media accounts that reported on public health concerns related to the pandemic. On the other hand, the *high bot score accounts*, of which only 0.05% is verified, may have used the COVID-19 conversation as a vector to promote conspiratorial narratives.

Lastly, we manually investigated the content of tweets in both groups discussed in the trigram analysis of original tweets. We document some findings next.[7] As for the *high bot score accounts*, the most popular tweets were posted by accounts that exhibit clear automation patterns; they have high friends/followers' ratio and posted thousands of tweets about COVID-19 following templated formats.

For example, we found hundreds tweets that contain references to news about COVID-19, followed by sequences of hashtags such as those displayed in **Figure 6A** in combination with a link, typically to Youtube videos (many of which had already been taken down by Youtube as of the time of this paper). Beside Youtube, the most referenced sources included various hyperpartizan news site, such as Infowars, ZeroHedge, etc. Typical sensationalistic headlines suggest that:

---

[7] Due to privacy requirements and in line with Twitter's Terms of Service, we here only refer to anonymized examples of tweets, removing any information that could be used for the reidentification of the authors.

(i)   the virus was made in Wuhan labs;
(ii)  the virus is a "globalist biological weapon";
(iii) the virus was imported into China by the US military;
(iv)  products imported from China may be infected with the virus.

Other hashtags and keywords associated with these narratives include free speech issues (#freespeech, #freezerohedge), "truthers" narratives (#coronavirustruth, #5G) allegedly uncovering globalist conspiracies, with a special emphasis on stories implying a connection between the diffusion of the 5G wireless technology, and the emergence of the virus originally in Wuhan, and a bot-fueled conspiracy ("5G Launches In Wuhan Weeks Before Coronavirus Outbreak") emerging in mid-February 2020.

As for *low bot score users,* they reference to traditional news sources, and the most referenced accounts are the US President, the CDC, the WHO and various established news organizations including both left and right leaning newspapers. As for their focus, in the earlier period (beginning of February) they tend to cover health-related generic hashtags such as #PublicHealth and #Health, as well as China-centric events keywords ("Wuhan", "China", etc.). Later in the observed period, their focus shifts on preventive measures (#WashYourHands, #StopTheSpread, #Quarantine) and lifestyle (#QuarantineLife); they also focus on policies and interventions enacted by the government (#SocialDistancing, #FlattenTheCurve); finally, they discuss the economy (#economy, #stockmarket, #bitcoin).

Concluding, the COVID-19 news bots tend to automatically collect and collate information from the news, and their most tweeted sources exhibit a mix of US-centric (the most popular being Fox News, the New York Times, CNN, and ABC News) and worldwide (e.g., The Guardian) news sources and organizations (e.g., Reuters) in English, as well as Twitter-based news bots (e.g., SmartNews). The typical tweet by news bot contains the headline of the news as body of the tweet text alongside with the URL. The large majority of the observed news bots tweets have been captured because the headlines contain one of the keywords used to seed the real-time data collection [19].

## DISCUSSION

COVID-19 is a global crisis and with people being pushed out of physical spaces due to containment measures, online conversation on social media becomes one of the primary tools to track social discussion. In fact, topics of conversation related to COVID-19 have been trending, uninterruptedly or so, ever since the beginning of the outbreak in early 2020. We leveraged a large-scale data collection tracking in real-time COVID-19 tweets since January 21, 2020, the day the first COVID-19 case was reported on US soil. The dataset we adopt here goes through March 12, 2020, the day before the United States government announced the state of national emergency due to the COVID-19 pandemic.

In this paper, we provided an early characterization of the prevalence of accounts that are likely automated and that post content in relation to the ongoing COVID-19 pandemic. To the best of our knowledge, this is the first study to provide some evidence that *high bot score accounts* are used to promote political conspiracies and divisive hashtags alongside with COVID-19 content.

## LIMITATIONS OF THIS STUDY

Our study has several limitations. First and foremost, despite the sheer size of the dataset at hand, we are only observing a small fraction, approximately 1%, of the overall Twitter conversation, through the lens of the Twitter API. This has been shown to introduce some biases toward over-represented topics [87], and COVID-19 has been the most spoken-about topic of discussion ever since the beginning of the pandemic. Second, another form of bias is automatically introduced when selecting keywords to follow. For example, despite our dataset exhibits dozens of languages, English content represents over two-third of the overall tweets. To mitigate this bias, we concentrated only on English content for this analysis, despite the fact that several interesting phenomena related to the scope of this work may be observed in other languages as well.

The third and most important limitation is related to the challenge of bot detection. Detecting bots is quite hard. Even the most sophisticated machine learning tools have varying levels of accuracy, especially when applied "in the wild" to identify bots in live conversations. To mitigate this issue, in this work we carried out three forms of additional validation, collecting data for suspended and verified accounts, and manually inspecting data of particular relevance or interest.

However, these solutions also have inherent limitations: for example, the list of suspended accounts only reflects Twitter's policies to intervene and ban an account but does not provide the rationale for the decision. Furthermore, the manual assessment can be viable for the scrutiny of few case studies, like those tackled in this paper, but is not a scalable strategy to carry out large scale studies. Nevertheless, our analysis (cf. Fig 1) illustrates that there is a strong correlation between increased bot score and higher probability of account suspension. Other key indicators such as the intensity of posting about COVID-19 as a function of account age can also be informative. This suggests that suspension algorithms employed by Twitter may benefit from accounting for such behavioral signatures to improve their accuracy.

## CONCLUSIONS

In this work, we set forth to investigate two research questions, namely whether automated social media accounts were active in the context of COVID-19 related discussion on Twitter (RQ1), and if so, whether they are engaged in bot behavior similar to what has been observed in prior work, e.g., fueling conspiratorial and ideological narratives (RQ2).

Our findings paint a picture where accounts that are likely automated have been used in malicious manners. We observed how *high bot score accounts* use COVID-19 as a vector to promote visibility of ideological hashtags that are typically associated with the alt-right in the United States. We have discussed the implications of this discovery and differentiated it from the behavior of human users that are predominantly concerned with public health and welfare.

In the future, we will further investigate the behavior of COVID-19 bots, to understand whether they are exclusively used for propaganda purposes, or whether other uses emerge. Preliminary analysis suggests that some bots may have been used as a form of citizen journalism to unearth information that would otherwise be censored in other platforms in China and surface it in the English-speaking Twitter [88].

Further evidence is needed to assess whether this is a systematic effort to promote freedom of speech. We will also delve deeper in the types of conspiratorial narratives that bots promote [89], and study the role of bots in the spread of rumors and fake news [90].

ACKNOWLEDGEMENTS

The author is grateful to the members of his lab for their invaluable help and support, in particular Emily Chen for initializing the tweet collection, Shen Yan for collecting the list of suspended accounts, and Ashok Deb for collecting the Botometer scores; the author is also grateful to Jason Baumgartner (PushShift.io) for sharing the list of verified Twitter accounts.

ABOUT THE AUTHOR

Emilio Ferrara is Research Assistant Professor of Computer Science at the University of Southern California and Research Team Leader at the USC Information Sciences Institute. His research focuses on characterizing information diffusion and manipulation of social media. He is a recipient of the 2016 Complex System Society Junior Scientific Award, and he received the 2016 DARPA Young Faculty Award.